\documentclass[a4paper]{jpconf}
\usepackage{graphicx}
\begin{document}
\title{Anti-nuclei and nuclei production in Pb+Pb collisions at CERN
SPS energies}

\author{V\,I Kolesnikov for the NA49 collaboration}
\address{Joint Institute for Nuclear Research, Dubna, Russia}
\ead{kolesnik@moonhe.jinr.ru}

\begin{abstract}
We present new results on production of $^3$He and $t$
obtained by the NA49 experiment in 20A, 30A, 40A and 80A GeV central
Pb+Pb collisions at the CERN SPS. Transverse mass spectra and rapidity
distributions for clusters measured over a large phase space domain are
discussed. We observe a weak dependence of the mid-rapidity $t$/$^3$He ratio on collision energy at SPS. The energy dependence of the total yield
for $^3$He is remarkably reproduced by a statistical hadron gas model.
A gradual decrease of the coalescence parameter B$_3$ for $^3$He
with $\sqrt {s_{NN}}$ is observed.
In addition, a measurement of the yield of anti-deuterons
in Pb+Pb reactions at SPS's top energy of $\sqrt {s_{NN}}$\,=\,17.2 GeV
is presented.
\end{abstract}

\vspace{-7mm}

\section{Introduction}
\hspace{3mm}
The aim of the measurements, performed by the NA49 experiment
during the energy scan program at the CERN SPS,
was to investigate the properties of strongly interacting matter
in heavy-ion collisions via a vast range of observables
from light hadrons ($\pi$, K) to light nuclear clusters ($^3$He, $t$).
Data on composite particle production
in reactions with heavy ions gives us valuable information about
the late stage of the fireball evolution and may provide a measure
of the size of the particle emitting source.
In addition, a comparison of experimental data on cluster yields
to the statistical model expectations can also shed some light on
the mechanism of formation of light nuclei in heavy ion collisions.

\section{Experiment NA49 and data analysis}
\hspace{3mm}
The main components of the NA49 apparatus \cite{na49} are four time
projection chambers (TPCs) (two of them are placed within magnetic fields produced by the superconducting magnets) for tracking and particle
identification (PID) via ionization energy loss dE/dx measurements.
The time-of-flight system (two TOF scintillator arrays situated beyond the
TPCs) provides timing information for PID covering the mid-rapidity region.
The downstream zero degree calorimeter was used for triggering and
centrality determination.
This report is based on the analysis of the data collected
during the 1999-2002 running period. A total of $1.2\cdot 10^6$
events representing the 7\% most central Pb+Pb collisions
at 20A, 30A, 40A and 80A GeV
were used in the study of $^3$He and $t$ production.
Anti-deuterons are measured in the 23\% most central Pb+Pb collisions
at 158A GeV ($2.6\cdot 10^6$ events).
Combined dE/dx and TOF information was used for the identification
of single charged hadrons and (anti)clusters at mid-rapidity.
Double charged $^3$He clusters are identified over almost
the entire phase space
via the dE/dx method.
The results presented below include corrections for the PID track quality
cuts, background contamination, detection efficiency and geometrical
acceptance. The detailed description of the analysis procedure is given
in Refs. \cite{na49_pd, na49_ppbar}.

\section{Results and discussion}

\begin{figure}[h]
\begin{minipage}{15pc}
\includegraphics[width=15pc]{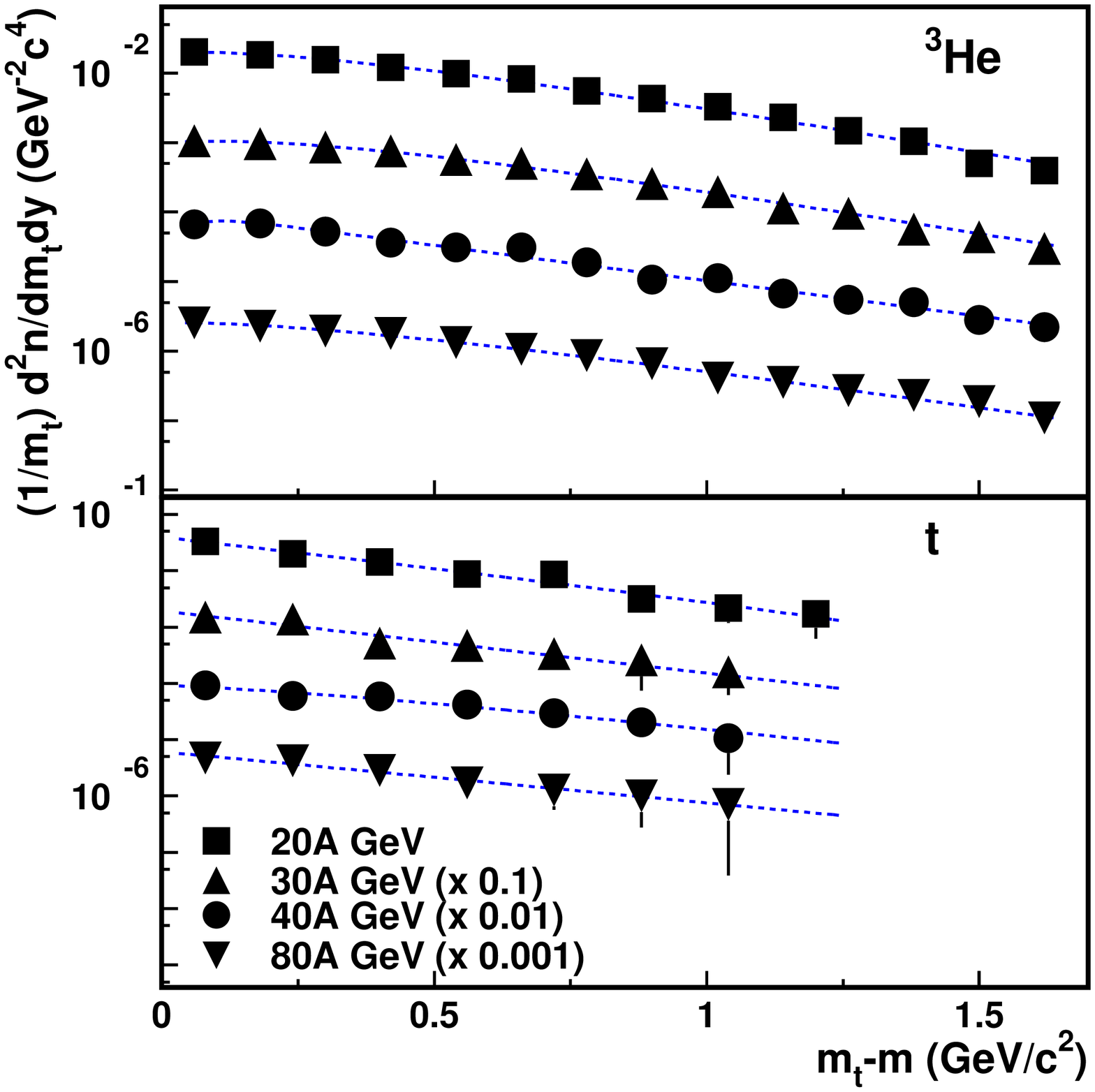}
\end{minipage}\hspace{3pc}%
\begin{minipage}{15pc}
\includegraphics[width=15pc]{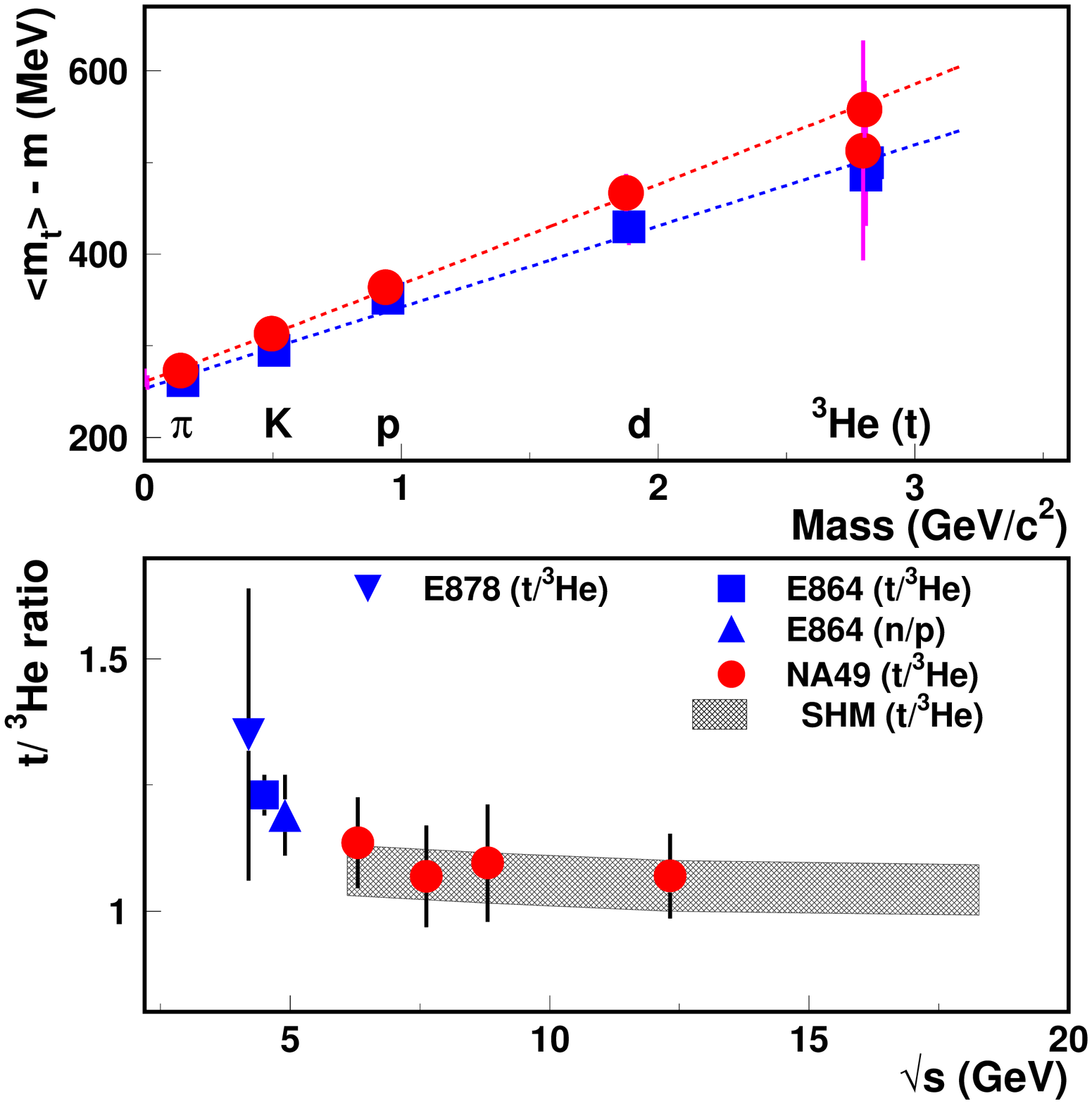}
\end{minipage}
\caption{\label{fig1}Left: the mid-rapidity m$_t$-spectra for $^3$He
(upper panel) and $t$ (lower panel) for the 7\% most central Pb+Pb
 collisions
(dashed lines show the double-exponential fits used for extrapolation
to the unmeasured range). Right: (upper panel) $<$$m_t$$>$-m versus particle
mass as obtained from the fits to the spectra at 20A(\fullsquare) and 80A GeV(\fullcircle); (lower panel)
mid-rapidity $t$ to $^3$He ratio as measured (points) and predicted
by the SHM model \cite{becat} (band).}
\end{figure}

\hspace{3mm}Fig.\,\ref{fig1} (left panel) shows the mid-rapidity transverse
mass spectra for helium-3 and tritons
observed in central Pb+Pb collisions at 20A-80A GeV.
As expected, the collective transverse flow flattens m$_t$-spectra for clusters at low m$_t$, so that the distributions shown are fitted with
a double exponential function (fits are shown as dashed lines in
Fig.\,\ref{fig1}).
The mean transverse mass values as obtained from the fits
at 20A and 80A GeV are plotted in Fig.\,\ref{fig1}
(right panel, upper plot) together with the NA49 measurements for hadrons
($\pi$, K, $p$) and deuterons \cite{na49_pd}.
Here, one can see a clear indication of a large collective transverse
flow effect: a considerable increase of $<$$m_t$$>$-$m$ with particle mass.
As Polleri {\it et al.} argued \cite{polleri}, a linear
dependence of mean $<$$m_t$$>$ on the particle mass futhermore supports
the conclusion that the particle emitting source has a uniform (box-like)
density distribution and a linear velocity profile.
There is no significant difference in resulting values of $<$$m_t$$>$
for particle species between 20A and 80A GeV data. This may indicate little
change in the strength of the transverse expansion over the SPS energy
domain.

\begin{figure}[h]
\begin{minipage}{12pc}
\includegraphics[width=12pc]{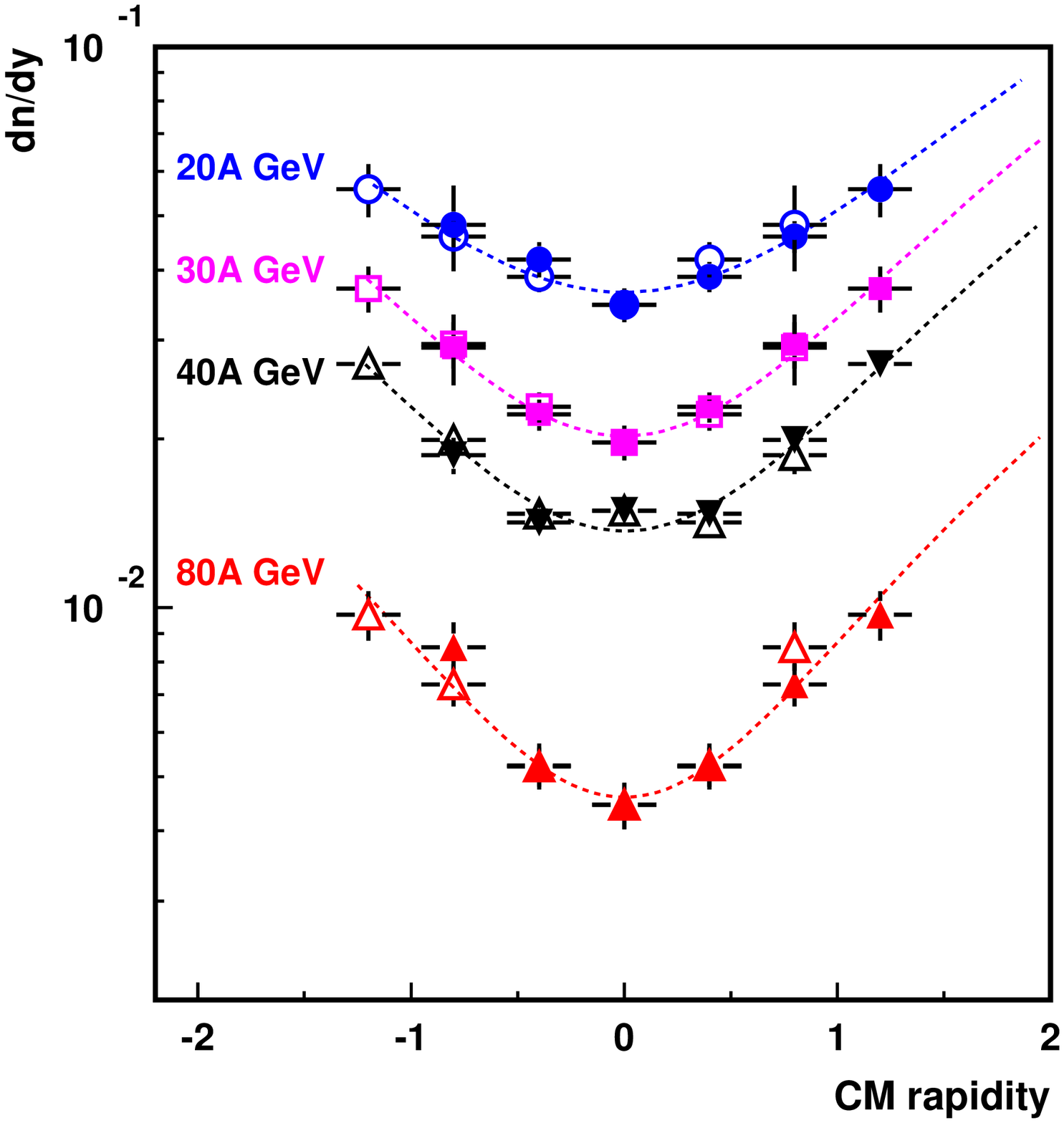}
\end{minipage}\hspace{1pc}
\begin{minipage}{12pc}
\includegraphics[width=12pc]{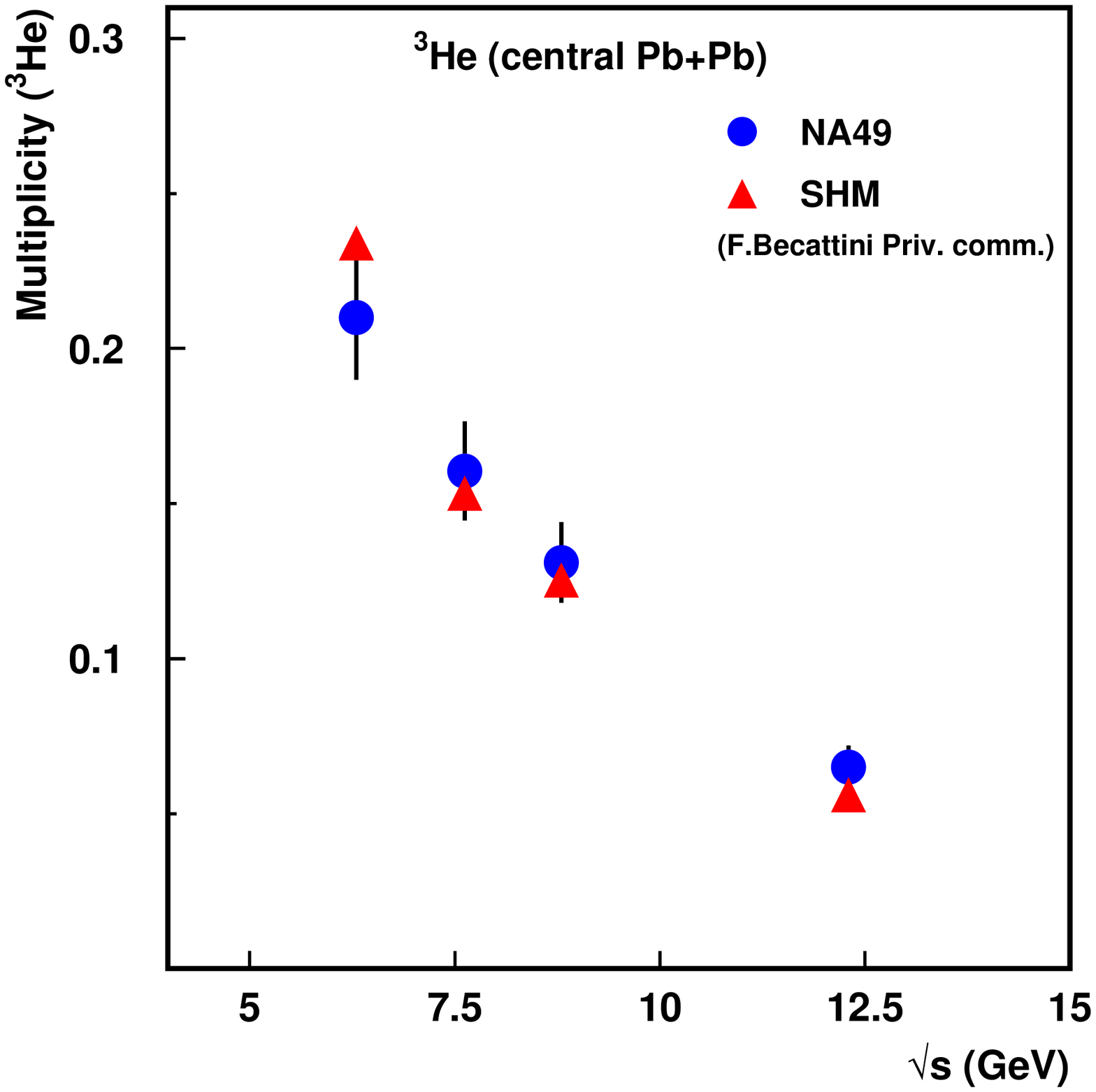}
\end{minipage}\hspace{1pc}
\begin{minipage}{12pc}
\vspace{2mm}
\includegraphics[width=11pc]{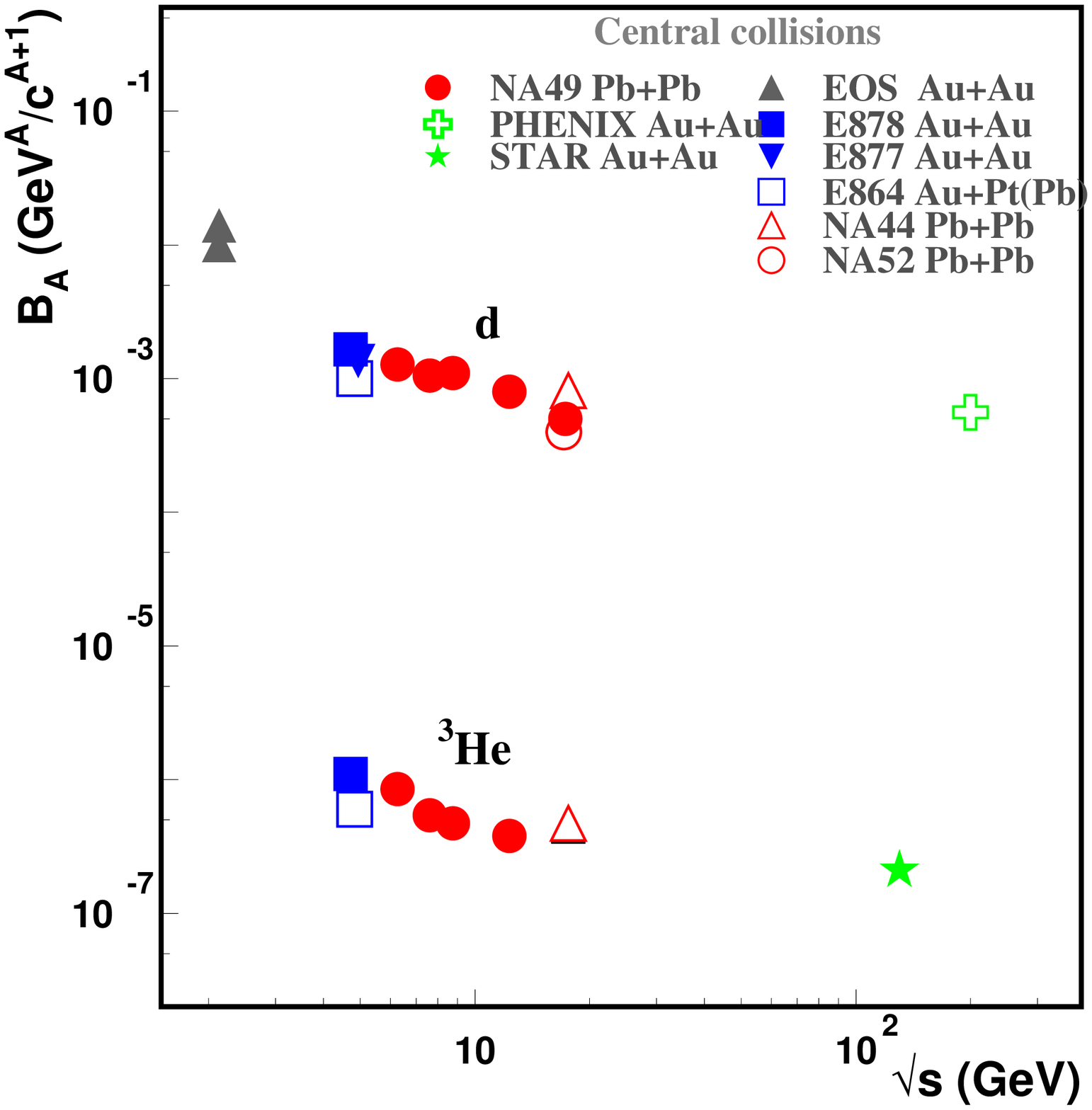}
\end{minipage}
\caption{\label{fig2}Left: the rapidity distributions and the parabolic
fits (dashed lines) for $^3$He in Pb+Pb at 20A-80A GeV
(open symbols are obtained by reflection at mid-rapidity).
Center: total yields of $^3$He as measured by NA49 (circles) and predicted
by the SHM model (triangles).
Right: energy dependence for
B$_2$ and B$_3$ in central A-A collisions.}
\end{figure}

For the complete picture of nuclear cluster production, information
about the phase space distribution of neutrons in the final state (at freezeout) is of importance. However the yield of $n$ (usually) remains
unmeasured.
As it has been established in the RQMD model,
the initial ratio of neutrons participating in the collision to protons
($n$:$p$=1.54:1 for $^{208}$Pb) changes considerably toward the equilibrium
value of $n$/$p$\,=\,1 during the fireball evolution as a result of strong
resonance production.
Assuming a simple additive scheme which relates the yield of the cluster
to the product of the yields of nucleons,
the freezeout $n$/$p$ ratio may be deduced from the
$t$/$^3$He ratio.
The right bottom panel of Fig.\,\ref{fig1}
shows the ratio $t$/$^3$He as a function of $\sqrt {s_{NN}}$. We observe a weak energy dependence of this ratio in central Pb+Pb collisions at
$\sqrt {s_{NN}}$$\,>\,$6 GeV. The average value $<$$t$/$^3$He$>$ is about 1.1 at SPS, which indicates a large degree of equilibrium in the
final state of the reaction in this energy domain.
This trend is well reproduced by the Statistical Hadronization Model (SHM)
\cite{becat} (SHM predictions are shown by the dark band in Fig.\,1).
The yields of $^3$He in Pb+Pb collisions at 20A-80A GeV extracted in
rapidity slices of $\Delta$y=0.4 are shown in Fig.\,\ref{fig2} (left) as
a function of rapidity. The rapidity distributions for $^3$He
are concave at all energies while those for protons are essentially flat
around mid-rapidity \cite{na49_qm06}.
The observed increase of $^3$He formation rate at very forward rapidities
in central Pb+Pb collisions has not been explained yet.
The total yields for $^3$He were obtained by fitting the measured
rapidity distributions with a parabola (fits are shown as dashed lines).
The 4$\pi$ yields of $^3$He are plotted in Fig.\,\ref{fig2} (center) as a function of $\sqrt {s_{NN}}$.
Also shown are the total multiplicities of $^3$He predicted by the  SHM model. The agreement with the NA49 measurements is remarkable.

\vspace{-5mm}

\begin{figure}[h]
\begin{minipage}{10.5pc}
\includegraphics[width=10.5pc]{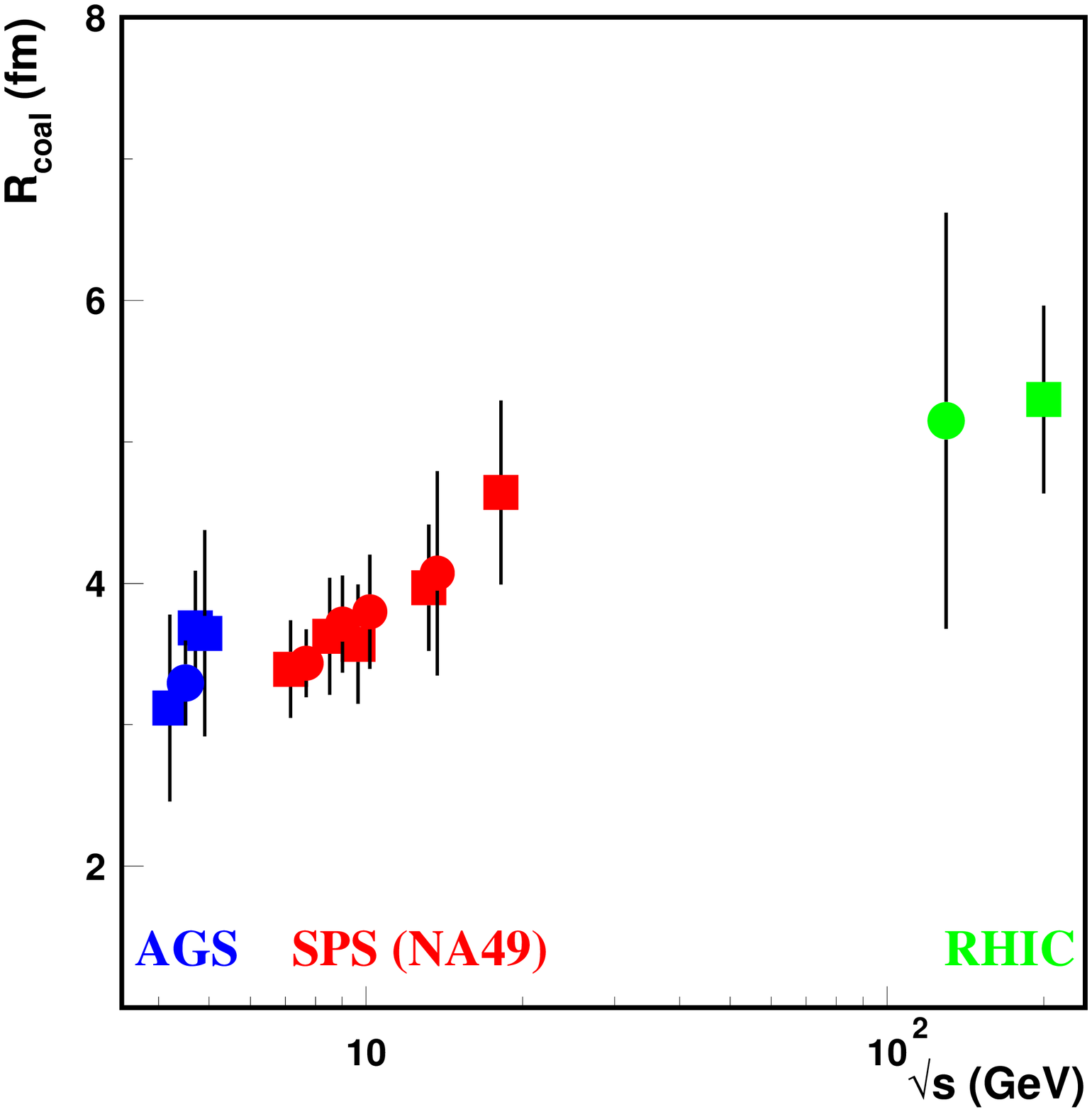}
\end{minipage}\hspace{1pc}%
\begin{minipage}{11.5pc}
\includegraphics[width=11.5pc]{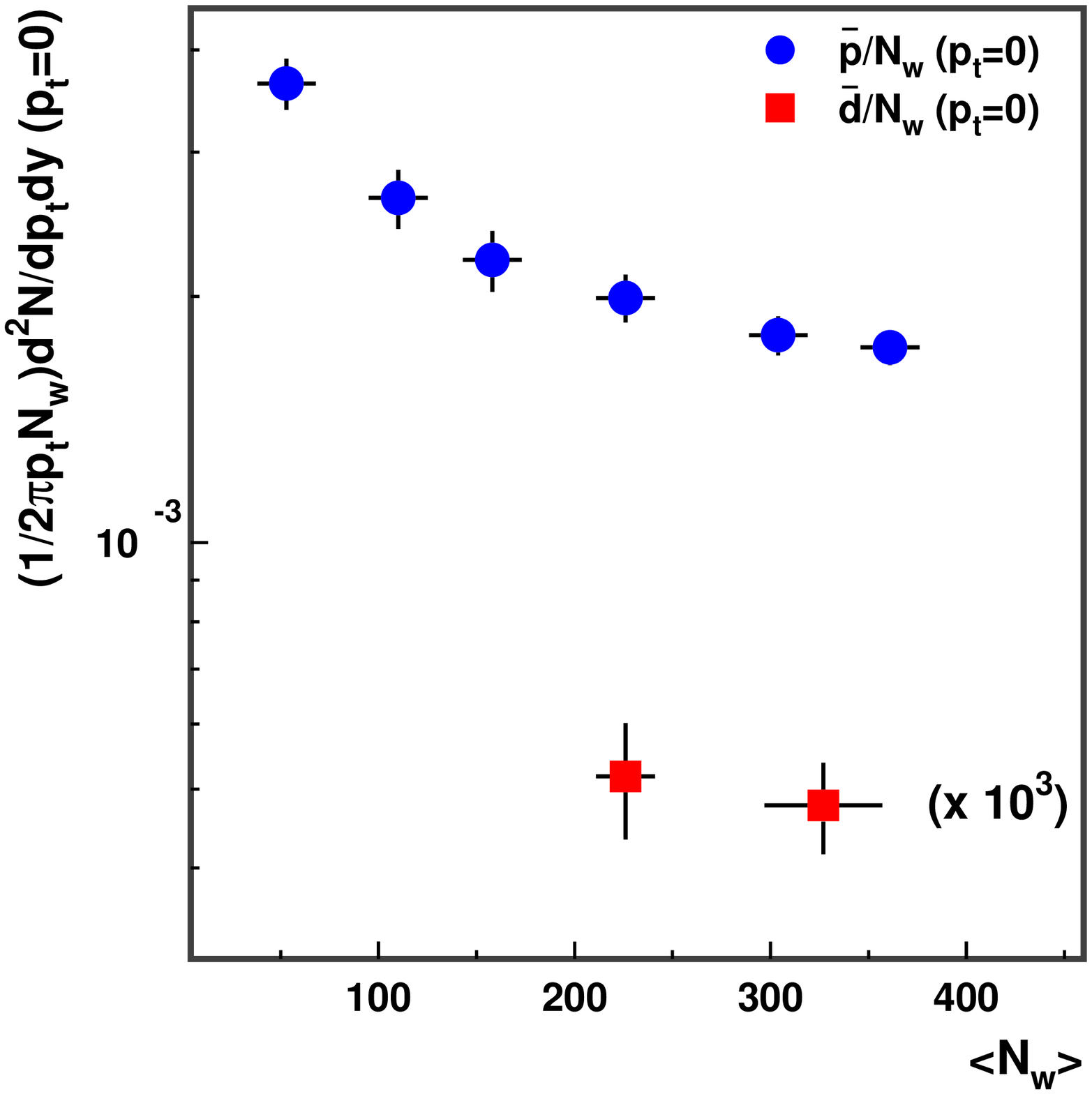}
\end{minipage}\hspace{1pc}%
\begin{minipage}{10.5pc}
\includegraphics[width=10.5pc]{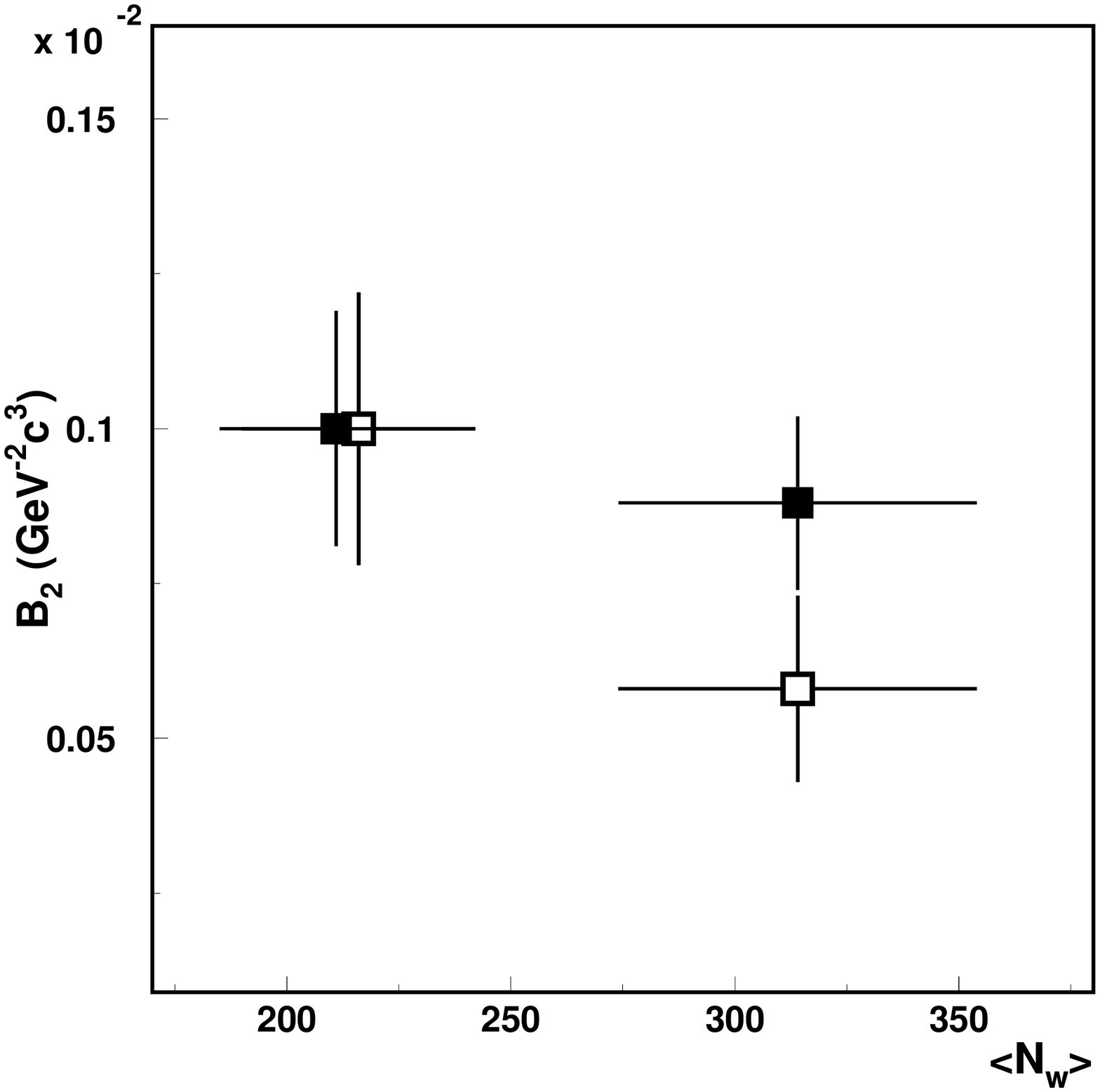}
\end{minipage}
\caption{\label{fig3}Left: R$_{coal}$ for $d$(\fullsquare)
and $^3$He(\fullcircle) in central A-A collisions at AGS (blue),
SPS (red) and RHIC (green). Center: invariant cross section per wounded
nucleon at $p_t$=0 as a function of $<$N$_w$$>$ for anti-protons
(\fullcircle) and anti-deuterons (\fullsquare). Right: B$_2$ for deuterons
(\opensquare) and anti-deuterons (\fullsquare) as a function of
 $<$N$_w$$>$.}
\end{figure}

A typical coalescence prescription \cite{coal1,coal2,coal3} relates the
invariant yield of
light nuclei of atomic mass number A to that of protons raised to the Ath
power ($n$ and $p$ distributions are assumed to be the same)
through a dimensioned variable - coalescence parameter B$_A$ as:
$$
E_A\frac{d^3N_A}{d^3P}=B_A\left(E_p\frac{d^3N_p}{d^3p}\right)^A, \hspace{1cm} P=A\cdot p
$$
B$_A$ can be converted, under specific assumption,
into the volume of the fireball at freeze-out (B$_A$ is inversely
related to that volume).
Fig.\,\ref{fig2} (right panel) shows energy dependence for coalescence parameters B$_2$ and B$_3$ in central heavy ion collisions.
Our measurements (circles) are plotted together with
 AGS\,\cite{b2_ags1,b2_ags2} and
RHIC\,\cite{b2_rhic1,b2_rhic2} data.
One can see, that both B$_3$ and B$_2$ decrease
as $\sqrt {s_{NN}}$ increases, suggesting increasing freeze-out volumes.
To examine this general trend futher, the (coalescence) radii
of the emitting source have been extracted,
using the prescription of Scheibl and Heinz \cite{heinz} for a
thermalized fireball with transverse flow.
A comparison of the obtained radii for $d$ and $^3$He with
those measured at other energies
is shown in the left panel of Fig.\,\ref{fig3}.
It is seen, that the source sizes for  different cluster species
agree with each other within the error bars and are found to be
rising with center-of-mass energy.

Now we turn to anti-deuterons.
In Fig.\,\ref{fig3} (center) the centrality dependence of the invariant
yield of anti-deuterons and anti-protons \cite{na49_ppbar} normalized to
the number of wounded nucleons is shown. Anti-deuterons are measured in
two centrality bins, corresponding to the 0-10\% and 10-23\% most central
Pb+Pb collisions. The yield of anti-deuterons per wounded nucleon
exhibits very weak variation with centrality
in the measured range in a manner similar to that observed for
anti-protons. B$_2$ for both deuterons and anti-deuterons, measured in
these event samples, is plotted
in Fig.\,\ref{fig3} (right panel). The B$_2$ values for deuterons agree
with those for anti-deuterons within the errors. The observed centrality
dependence suggests increase of the source size in more central collisions.

\section{Summary}
\hspace{3mm}The NA49 experiment has measured $^3$He and $t$ production
 in central
Pb+Pb collisions at 20A-80A GeV. The invariant yields for clusters
are described by a sum of two exponential functions in $m_t$ and the $<$$m_t$$>$ values appear to follow a linear increase with particle mass.
The mid-rapidity $t$/$^3$He ratio in central Pb+Pb collisions at SPS
energies is measured to be $t$/$^3$He\,$\approx$\,1.1, which is
considerably smaller than the initial participant's n/p ratio of 1.54.
We observe that the rapidity distributions for $^3$He are concave
at all studied energies.
It appears that a statistical hadron gas model is able to reproduce data
on $^3$He yields.
B$_3$  and B$_2$ coalescence parameters follow a decreasing trend with
collision energy.
The source radii deduced from the measured B$_2$ and B$_3$ parameters
are found to be consistent.

\ack{
This work was supported by the US Department of Energy Grant
DE-FG0397ER41020/A000, the Bundesministerium fur Bildung und Forschung,
Germany, the Virtual Institute VI-146 of Helmholtz Gemeinschaft, Germany,
the Polish State Committee for Scientific Research (1 P03B 006 30,
1 P03B 097 29, 1 P03B 127 30), the Hungarian Science Research Foundation
(T032648, T032293, T043514), the Hungarian National Science Foundation,
OTKA, (F034707), the Polish-German Foundation, the Korea Science \& Engineering
 Foundation (R01-2005-000-10334-0) and the Bulgarian national Science Fund
 (Ph-09/05)}

\end{document}